\documentclass[trackchanges,twocolumn]{aastex701}

\usepackage{siunitx}

\begin{document}

\title{Sloshing Motions in Abell 3571 Revealed by XRISM/Resolve Velocity Mapping}

\author[orcid=0009-0009-8380-1260,gname=Itsuki,sname=Aihara]{Itsuki~Aihara}
\affiliation{Department of Physics, Tokyo University of Science, 1-3 Kagurazaka, Shinjuku-ku, Tokyo 162-8601, Japan}
\email[show]{itsukiaihara0726@gmail.com}

\author[orcid=0000-0001-5888-7052,gname=Congyao,sname=Zhang]{Congyao~Zhang}
\affiliation{Masaryk University, Department of Theoretical Physics and Astrophysics}
\affiliation{Department of Astronomy \& Astrophysics, University of Chicago, Chicago, IL 60637 USA}
\email{cyzhang@astro.uchicago.edu}

\author[gname=Sora,sname=Nakajima]{Sora~Nakajima}
\affiliation{Department of Physics, Tokyo University of Science, 1-3 Kagurazaka, Shinjuku-ku, Tokyo 162-8601, Japan}
\email{1225552@ed.tus.ac.jp}

\author[orcid=0000-0003-2907-0902,gname=Kyoko,sname=Matsushita]{Kyoko~Matsushita}
\affiliation{Department of Physics, Tokyo University of Science, 1-3 Kagurazaka, Shinjuku-ku, Tokyo 162-8601, Japan}
\email[show]{matusita@rs.tus.ac.jp}

\author[orcid=0000-0003-3537-3491]{Hannah~McCall}
\affiliation{Department of Astronomy \& Astrophysics, University of Chicago, Chicago, IL 60637 USA}
\email{hannahmccall@uchicago.edu}

\author[orcid=0000-0001-7630-8085]{Irina~Zhuravleva}
\affiliation{Department of Astronomy \& Astrophysics, University of Chicago, Chicago, IL 60637 USA}
\email{zhuravleva@uchicago.edu}

\author[orcid=0000-0001-7773-9266,gname=Shogo, sname=Kobayashi]{Shogo~B.~Kobayashi}
\affiliation{Rikkyo University, Department of Physics, 3-34-1 Nishi Ikebukuro, Toshima-ku, Tokyo 171-8501, Japan}
\email{shogo.kobayashi@rikkyo.ac.jp}

\author[orcid=0000-0001-8055-7113, gname=Kotaro, sname=Fukushima]{Kotaro~Fukushima}
\affiliation{Department of Physics, Tokyo University of Science, 1-3 Kagurazaka, Shinjuku-ku, Tokyo 162-8601, Japan}
\email{kxfukushima@gmail.com}

\author[orcid=0000-0002-9478-1682, gname=William, sname=Forman]{William~Forman}
\affiliation{Harvard-Smithsonian Center for Astrophysics, 60 Garden St., Cambridge, MA, 02138 USA}
\email{wforman@cfa.harvard.edu}

\author[orcid=0000-0003-2206-4243, gname=Christine, sname=Jones]{Christine~Jones}
\affiliation{Harvard-Smithsonian Center for Astrophysics, 60 Garden St., Cambridge, MA, 02138 USA}
\email{cjones@cfa.harvard.edu}

\author[orcid=0000-0003-3537-3491]{Annie~Heinrich}
\affiliation{Department of Astronomy \& Astrophysics, University of Chicago, Chicago, IL 60637 USA}
\email{amheinrich@uchicago.edu}

\author[orcid=0000-0003-3537-3491]{Daniele~Rogantini}
\affiliation{Department of Astronomy \& Astrophysics, University of Chicago, Chicago, IL 60637 USA}
\email{danieler@uchicago.edu}

\author[orcid=0000-0001-5774-1633, gname=Kosuke, sname=Sato]{Kosuke~Sato}
\affiliation{Department of Astrophysics and Atmospheric Sciences, Kyoto Sangyo University, Motoyama, Kamigamo, Kita-ku, Kyoto, Kyoto 603-8555, Japan}
\email{ksksato@cc.kyoto-su.ac.jp}

\author[gname=Kazunori,sname=Suda]{Kazunori~Suda}
\affiliation{Department of Physics, Tokyo University of Science, 1-3 Kagurazaka, Shinjuku-ku, Tokyo 162-8601, Japan}
\email{1225703@ed.tus.ac.jp}

\author[orcid=0000-0003-3701-5882]{Ildar~Khabibullin}
\affiliation{Rudolf Peierls Centre for Theoretical Physics, Department of Physics, University of Oxford, Clarendon Laboratory, Parks Rd, Oxford, OX1 3PU, United Kingdom}
\affiliation{Max Planck Institut f\"ur Astrophysik, Karl-Schwarzschildstr.~1, D-85748 Garching, Germany}
\email{ildar@mpa-garching.mpg.de}

\begin{abstract}
    Minor mergers can induce sloshing motions in the intracluster medium, leaving characteristic signatures in the thermodynamic structure and gas kinematics of cluster cores.
    Abell~3571 is an X-ray-bright, apparently relaxed cluster at $z \sim 0.04$.
    We observed the central $\sim 300$ kpc region of Abell~3571 with four partially overlapping XRISM Resolve pointings, 
    covering three contiguous Resolve fields to the north, south and east with a total exposure time of approximately 575 ks.
    The velocity dispersions are subsonic and are at the level of $\sim 100$--$150 ~\mathrm{km~s^{-1}}$ across most regions.
    The cooler region associated with the northern surface-brightness excess is blueshifted by up to $\sim -60 ~ \mathrm{km~s^{-1}}$ relative to the brightest cluster galaxy (BCG),
    while the hotter region in the southern and eastern surface-brightness deficit regions is redshifted by up to $\sim 170 ~ \mathrm{km~s^{-1}}$.
    Numerical simulations suggest that this large-scale thermodynamic and kinematic asymmetry is broadly consistent with early-phase sloshing induced by an off-axis minor merger.
    Abell~3572, an X-ray-faint gas-poor cluster located 1.6~Mpc to the south,
    is a promising candidate for the perturber.
    Given the lack of clear signatures of prominent AGN feedback in Abell~3571, 
    these results suggest that sloshing-driven gas redistribution may contribute to delaying the re-establishment of a strong cool core in Abell~3571.
\end{abstract}

\keywords{\uat{Galaxy clusters}{584} --- \uat{Intracluster medium}{858} --- \uat{High Energy astrophysics}{739} --- \uat{Abell clusters}{9}}

\section{Introduction}\label{sec:intro}
    Galaxy clusters are the largest gravitationally bound structures in the Universe,
    composed mainly of dark matter and the hot intracluster medium (ICM). 
    They grow through mergers and continuous accretion, which drive gas motions efficiently within the ICM \citep[e.g.,][]{vazza2009, Miniati2015}.
    \textit{Chandra} observations have shown that cold fronts are common in cool-core cluster cores, suggesting that low-entropy central gas undergoes sloshing motions \citep[e.g.,][]{Markevitch2003}.
    Even a minor merger can excite such long-lived kinematic imprints in cluster cores \citep{Ascasibar2006, ZuHone2010, ZuHone2016}.
    Such motions transfer angular momentum to the cluster atmosphere which develops spiral-like structures. 
    These sloshing motions may contribute to the thermal balance of the core by transporting and dissipating energy \citep[e.g.]{ZuHone2010, ZuHone2016}, 
    in addition to heating by the active galactic nucleus (AGN) in the central galaxy \citep{McNamara2007}.
    Such gas motions, together with AGN feedback,
    are important for understanding the dynamical and thermodynamic evolution of galaxy clusters,
    including a possible transition between cool-core and non-cool core states \citep{Lehle2025}.

    The sloshing spirals are manifest as cold fronts --- sharp discontinuities in X-ray surface brightness and temperature, which have been identified in high-resolution X-ray images (e.g., from Chandra and XMM; see \citealt{Churazov2003, Simionescu2012}).
    However, direct measurements of gas kinematics have been challenging given the spectral resolution of CCD detectors, leaving uncertainties of approximately $\pm 200~\mathrm{km~s^{-1}}$ \citep{Sanders2020}.

    The XRISM satellite has marked a breakthrough since its launch in 2023, providing an energy resolution of $\sim 4.5~\mathrm{eV}$ at 6 keV with the onboard Resolve microcalorimeter \citep{Tashiro2025}.
    XRISM has now observed several galaxy clusters that span a range of dynamical states, and the results obtained so far suggest that relaxed or weakly disturbed systems generally show modest velocity dispersions \citep{XRISM2025_cen, Fujita2025, Rose2025, XRISM2026_M87}, whereas merging systems can exhibit larger bulk velocities and sometimes enhanced velocity dispersion\citep{XRISM2025_A2319, Omiya2026_A754, Heinrich2026}.
    Among these, XRISM observations of the Perseus cluster, the brightest cool-core cluster in the X-ray sky, provided a kinematic map based on multiple discrete pointings sampling several radial directions \citep{XRISM2025_per, Zhang2026}. 
    These observations revealed enhanced velocity dispersion around the very central region dominated by AGN feedback, as well as velocity structures on large scales consistent with sloshing simulations.
    Another notable example is M87, the central galaxy of the Virgo cluster. XRISM observations revealed strongly enhanced gas motions within the innermost $\sim 5~\mathrm{kpc}$, where the velocity field is dominated by AGN-driven activity, reaching Mach numbers of up to $\sim 0.7$ \citep{XRISM2026_M87}.
   
    Abell~3571 (A3571) is the sixth brightest galaxy cluster in X-rays and is located at a redshift of $z \sim 0.04$.
    Optical and X-ray images suggested that it is a relatively relaxed cluster \citep{Lopes2018}.
    The azimuthally averaged temperature and iron abundance profiles outside $\sim 230~\mathrm{kpc}$ are similar to those of the Perseus cluster, while A3571 does not have a prominent cool core \citep{Matsushita2011}.
    \cite{Rossetti2010} classified this cluster as a ``merger cool-core remnant'' because the brightest cluster galaxy (BCG) lies in a region of low entropy and high metallicity, while the outer core appears dynamically relaxed.
    Observations with the Einstein Probe have also revealed a surface brightness excess to the north and southwest.
    The galaxy density distribution and X-ray morphology suggest that A3571 is still recovering from a minor merger and is currently in a post-merger stage \citep{Venturi2008, Zheng2026}.
    Previous X-ray and radio observations do not show clear evidence for prominent AGN feedback in A3571, although possible cavities and a faint central radio source have been reported \citep{Venturi2002,Birzan2012,Olivares2023}.
    A3571 therefore is a promising target for investigating sloshing-induced gas motions and their role in the thermal and dynamical evolution of the ICM in an apparently relaxed cluster without a prominent cool core.
    
    In this letter, we present XRISM/Resolve observations of A3571, with particular focus on its dynamical state, especially bulk motions associated with gas sloshing.
    A companion paper from our team focuses specifically on the central 100 kpc of A3571 and presents a detailed analysis of resonant scattering, non-thermal pressure support, and comparisons with the cores of other galaxy clusters observed with XRISM \citep{McCall2026}.
    Throughout this Letter, we assume a flat cosmology with $H_0 = 70 ~ \mathrm{km~s^{-1}~Mpc^{-1}}$ and $\Omega_m = 0.27$
    At the redshift of A3571, \ang{;1;} corresponds to 49~kpc.
    Unless otherwise stated, all uncertainties are quoted at the confidence level $1 \sigma$.

    \begin{figure*}[ht!]
        \includegraphics[width=0.33\linewidth]{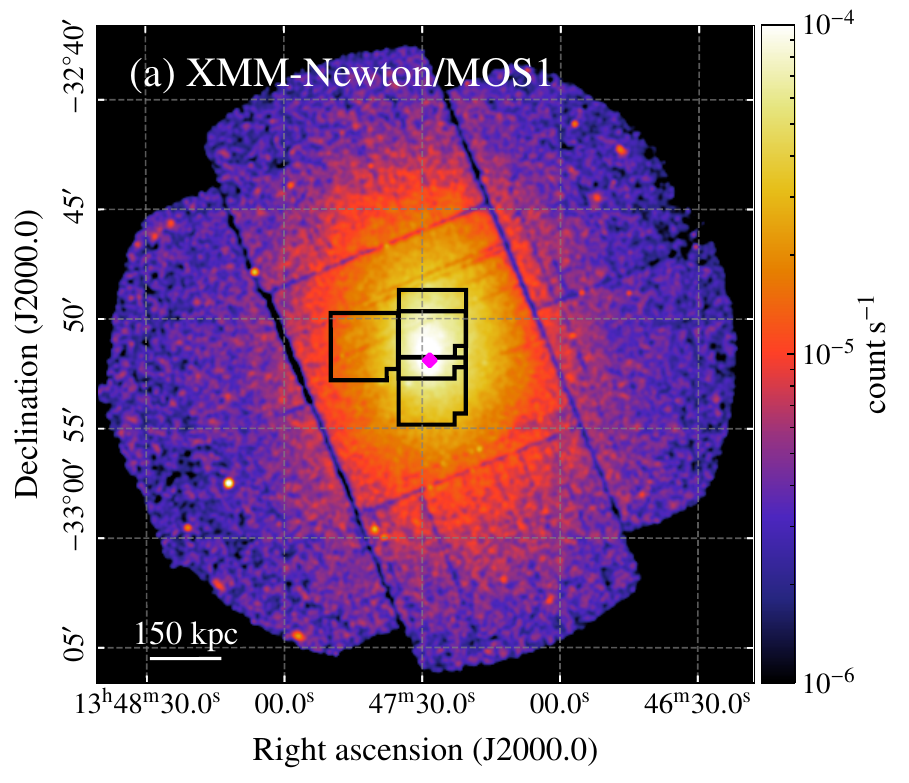}
        \includegraphics[width=0.33\linewidth]{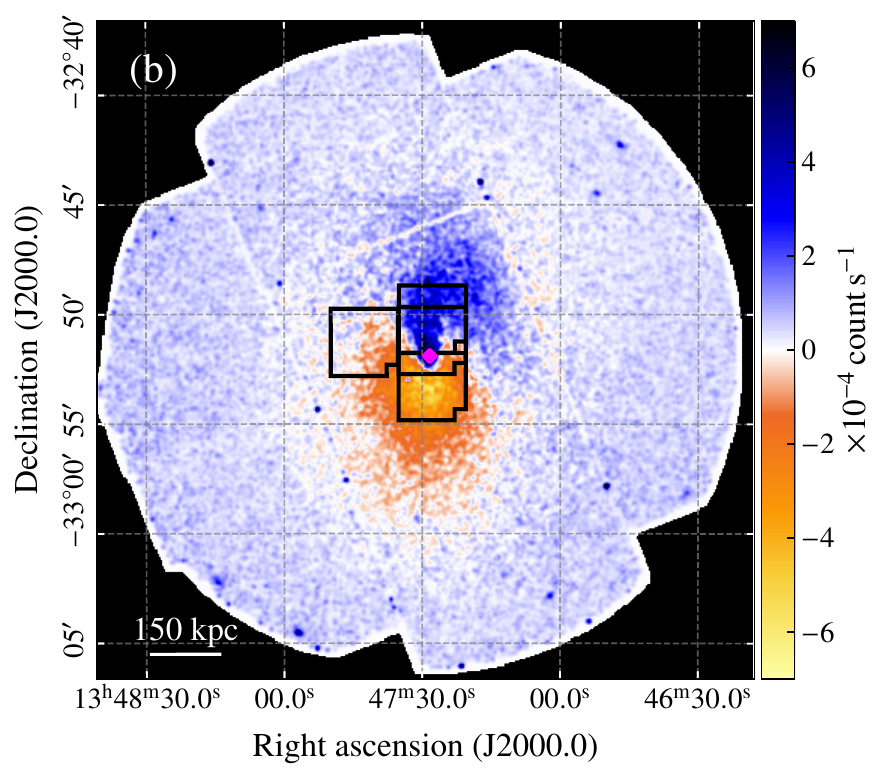}
        \includegraphics[width=0.33\linewidth]{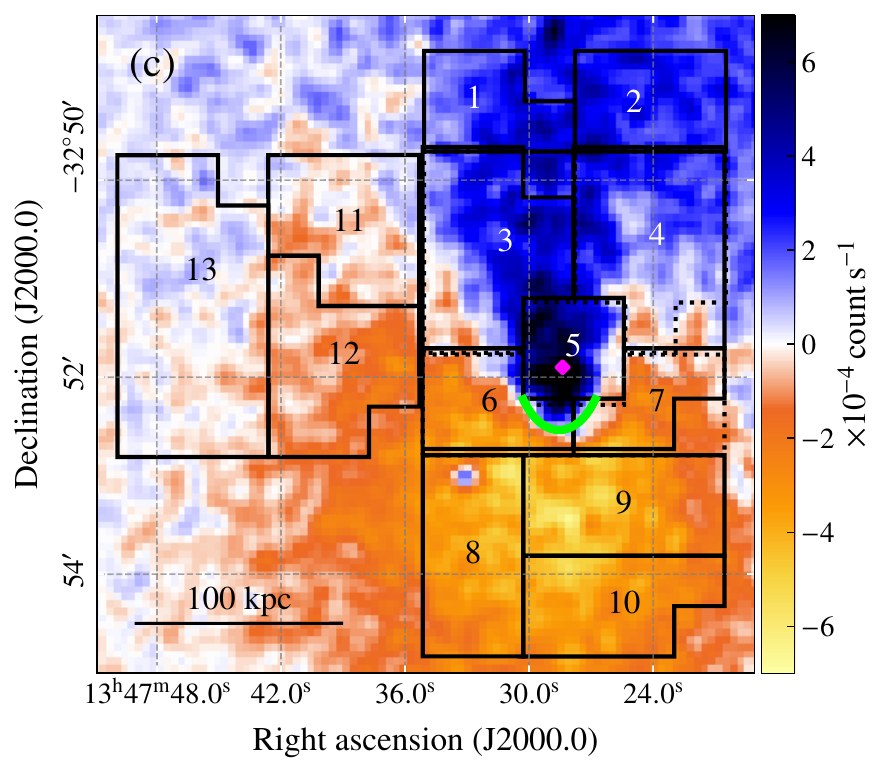}
        \caption{
            (a) Exposure-corrected X-ray image of A3571 obtained with XMM-Newton/MOS1 in the $0.5 \text{--} 2.0 \, \mathrm{keV}$ band.
            No background was subtracted, and the image was smoothed with a 2-dimensional Gaussian of $\sigma = 5.0 \, \mathrm{pixels}$ = \ang{;;5.5}.
            Black regions represent the Resolve FOV.
            (b) The residual image obtained by subtracting the best-fit elliptical $\beta$-model form the XMM/MOS1 image shown in the panel (a).
            Blue regions correspond to positive residuals, while red regions indicate surface brightness deficits.
            Black regions represent the Resolve FOV.
            (c) The same as (b),  shown at a larger scale.
            The black lines indicate the Resolve spectral extraction regions. 
            The dotted lines indicate the regions covered by two Resolve observation, i.e., by ABELL3571\_C and either ABELL3571\_S or ABELL3571\_N.
            The green curve highlights the surface brightness jump.
            The magenta diamond marks the position of the BCG in all panels.
            \label{fig:A3571img}
        }
    \end{figure*}
    
\section{Observation and Data Reduction}\label{sec:obs&reduction}
    \subsection{XRISM/Resolve Data}\label{subsec:XRISMdata}
        XRISM performed four partially overlapping observations of the central region of A3571 between 2024 December 30 and 2025 January 9.
        As shown in Figure~\ref{fig:A3571img} (a), the observations cover the core region (ABELL3571\_C; Obs.ID: 201095010, $13^\mathrm{h} 47^\mathrm{m} 27.85^\mathrm{s}$, \ang{-32;51;11.84}) and three offset pointings: \ang{;2;} south (ABELL3571\_S:Obs.ID: 201023010, $13^\mathrm{h} 47^\mathrm{m} 27.84^\mathrm{s}$, \ang{-32;53;18.24}), \ang{;1;} north (ABELL3571\_N; Obs.ID: 201024010, $13^\mathrm{h} 47^\mathrm{m} 27.78^\mathrm{s}$, \ang{-32;50;13.24}), and \ang{;3;} east (ABELL3571\_E; Obs.ID: 201025010, $13^\mathrm{h} 47^\mathrm{m} 42.60^\mathrm{s}$, \ang{-32;51;16.70}).
        
        To remove frame events from the cleaned event files of Resolve, we applied a screening condition of $46 < \text{RISE\_TIME+0.00075*DERIV\_MAX} < 58$.
        In this study, only high-resolution primary events (ITYPE = 0) are used.
        Pixel 27 was excluded from the analysis because its gain variation characteristics differ significantly from those of the other pixels.
        After screening, the effective exposure times are 192 ks (ABELL3571\_C), 175 ks (ABELL3571\_S), 137 ks (ABELL3571\_N), and 72 ks (ABELL3571\_E).
        
        The response files were generated using HEASoft v6.36 and the XRISM calibration database (CALDB) released on 2025-09-15.
        The redistribution matrix files (RMFs) were produced with \texttt{rslrmf}, adopting whichrmf$=$X.
        The ancillary response function (ARF) files were generated with \texttt{xaarfgen} in the IMAGE mode, using a count image obtained with XMM-Newton/MOS1 as the input image.
        We did not generate non–X-ray background (NXB) spectra with \texttt{rslnxbgen} from night-Earth data.

    \subsection{XMM-Newton Data}\label{subsec:XMMdata}
        XMM-Newton observed the central region of A3571 on 2002 July 29 (ObsID 0086950201, $13^\mathrm{h} 47^\mathrm{m} 28.90^\mathrm{s}$, \ang{-32;51;57.00}).
        We used only MOS1 data for the image analysis, because it provided sufficient counts per pixel for constructing a residual map for comparison, while the pn image was affected by CCD gaps that complicate the analysis. 
        Data screening was performed using the Science Analysis System (SAS) version 22.1.0.
        First, to remove poor-quality data and background events associated with incomplete CCD readout, we applied standard event screening with the \texttt{evselect} task.
        Next, periods affected by solar flares were excluded using the \texttt{tabgtigen} task.
        The filtering threshold was set to $0.22 ~ \mathrm{count~s^{-1}}$ corresponding to the count rates during quiescent periods.
        After screening, the effective exposure time was 26.6 ks.
        Using the screened data, we generated images in the 0.5--2.0 keV and 2.0--8.0 keV bands. The former was used for the imaging analysis presented in this work, while the latter was used as the input image for generating the XRISM/Resolve ARFs.
        
\section{Data Analysis and Results}\label{sec:analysis&result}
    \subsection{XMM-Newton Image Analysis}\label{subsec:XMManalysis}
        The exposure-corrected MOS1 image in the 0.5--2.0 keV band, shown in Figure~\ref{fig:A3571img}(a), 
        displays a fairly regular elliptical morphology elongated along the north--south direction.
        We fitted the image with an elliptical $\beta$ model, and subtracted the
        best-fit model to produce the residual map shown in Figure~\ref{fig:A3571img}(b).
        The residual map reveals a surface brightness excess extending northward from the BCG and bending toward the west. 
        A brightness jump is also visible about \ang{;;30} south of the BCG, while negative residuals appear farther south. 

    \subsection{Resolve Extraction Regions and SSM Analysis}\label{subsec:SSManalysis}
        The Resolve spectral extraction regions were defined from the XMM/MOS1 surface brightness residual map to trace the characteristic excess and deficit structures,
        as shown in Figure~\ref{fig:A3571img}(c).
        For diffuse sources, the broadening of the point spread function (PSF) produces non-negligible contamination from adjacent regions.
        To account for this effect and investigate the spatial structure of the ICM,
        we performed a spatial–spectral mixing (SSM) analysis.
        The fraction of photon leakage between regions was evaluated using \texttt{xrtraytrace} by simulating photon paths from each sky region to the detector,
        and ARFs were generated accordingly.
        In the ray-tracing simulation, the spatial distribution of the source photons was assumed to follow the 2.0--8.0 keV band image obtained with XMM-Newton EPIC/MOS1.
        Using these ARFs, we simultaneously fitted the 2--15 keV spectra of all regions,
        with the ICM components weighted according to the leakage fractions between the regions.
        For each region,
        the contributions of other regions were neglected when the leakage effective area was less than 5\% of that of the target region.
    
    \subsection{Resolve spectral analysis modeling}\label{subsec:XRISMmodel}
        \begin{figure*}[ht!]
            \plottwo{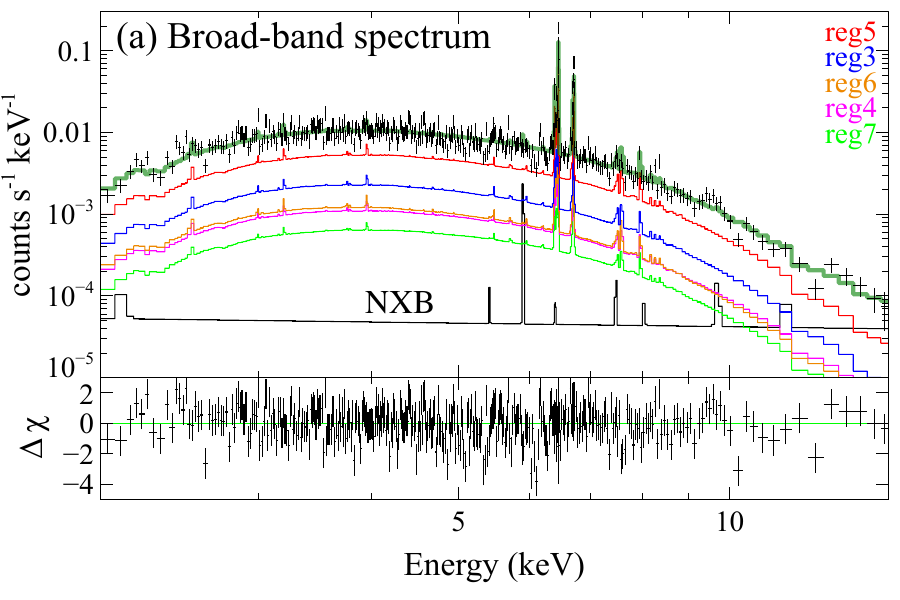}{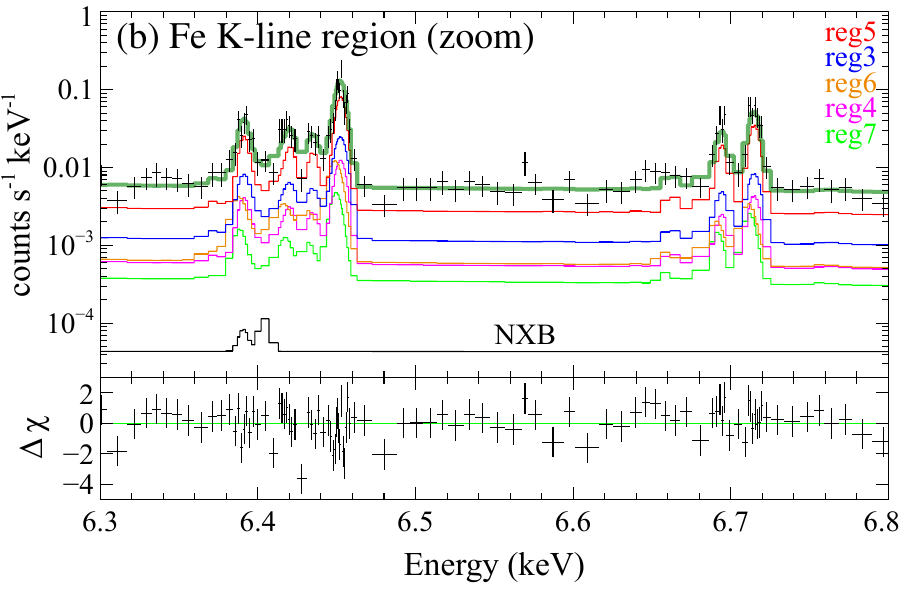}
            \caption{
                Representative spectrum and best-fit SSM model for region 5, which includes the BCG of A3571. The spectrum has been rebinned for display purposes only.
                The red model component represents the contribution from region 5 itself, while the other colored lines show the contributions of photons originating  in the neighboring regions.
                The black model component represents the NXB, and the thick green line represents the sum of all model components.
                The right panel shows a zoomed-in view of the Fe-line region of the same spectrum.
                \label{fig:exspec}
            }
        \end{figure*}
   
        Resolve spectral fitting was performed with XSPEC version 12.15.1 \citep{Arnaud1996}, adopting the C-statistic\citep{Cash1979}. 
        Elemental abundances were measured relative to the proto-solar values reported by \citet{Lodders2009}.
        The ICM was modeled with a collisionally ionized plasma model that included velocity broadening (\texttt{bapec}) \citep{Smith2001}, based on AtomDB version 3.1.3. 
        The free parameters were temperature ($kT$), abundance, velocity dispersion ($\sigma_\mathrm{los}$), redshift, and normalization. For regions covered by two observations, 
        indicated by dashed lines in Figure~\ref{fig:A3571img}(c), the model parameters were tied across observations except for normalization. The best-fit normalizations were broadly consistent between the two observations for the overlapping regions.
        Galactic absorption was modeled using the Tuebingen–Boulder interstellar medium absorption model (\texttt{tbabs}), with the hydrogen column density fixed at the Galactic value of $3.9 \times 10^{20} ~ \mathrm{cm^{-2}}$ 
        \citep{HI4PI2016} at the center of the Resolve field of view for the observation with Obs. ID 201095010.
        The NXB was modeled following the prescription provided by the XRISM team\footnote{\url{https://heasarc.gsfc.nasa.gov/docs/xrism/analysis/nxb/nxb\_spectral\_models.html}},
        using a phenomenological model consisting of a power-law component, three scale factors,and seventeen Gaussian emission lines that reproduce instrumental fluorescence features. 
        For the NXB model, the free parameters were the photon index of the power-law component, the line width of the Au emission line, and a constant factor.

        \begin{deluxetable*}{rrrrrr}
            \digitalasset
            \tablewidth{0pt}
            \tablecaption{Best-fit parameters of SSM Analysis.\label{tab:params}}
            \tablehead{
                \colhead{Reg. Num} & \colhead{$kT$} & \colhead{Abundance}\tablenotemark{a} & \colhead{$\sigma_\mathrm{los}$} & \colhead{Redshift} & \colhead{$u_\mathrm{los}$}\tablenotemark{b} \\
                \colhead{ }& \colhead{($\mathrm{keV}$)} & \colhead{($\mathrm{solar}$)} & \colhead{($\mathrm{km \, s^{-1}}$)} & \colhead{($\times 10^{-2}$)} & \colhead{($\mathrm{km \, s^{-1}}$)}
                }
            \startdata
                1 & ${6.4}^{+0.5}_{-0.4}$ & ${0.48}^{+0.08}_{-0.07}$ & ${55}^{+75}_{-55}$ & ${3.860}^{+0.007}_{-0.009}$ & ${31}^{+19}_{-25}$ \\
                2 & ${5.9}^{+0.4}_{-0.3}$ & ${0.53}^{+0.07}_{-0.06}$ & ${87}^{+34}_{-52}$ & ${3.827}^{+0.009}_{-0.006}$ & ${-63}^{+24}_{-17}$ \\
                3 & ${6.3}^{+0.2}_{-0.2}$ & ${0.51}^{+0.03}_{-0.03}$ & ${109}^{+16}_{-20}$ & ${3.841}^{+0.004}_{-0.003}$ & ${-22}^{+12}_{-10}$ \\
                4 & ${6.6}^{+0.2}_{-0.3}$ & ${0.53}^{+0.04}_{-0.04}$ & ${87}^{+21}_{-22}$ & ${3.841}^{+0.005}_{-0.005}$ & ${-24}^{+13}_{-13}$ \\
                5 & ${6.9}^{+0.9}_{-0.4}$ & ${0.64}^{+0.18}_{-0.09}$ & ${0}^{+60}_{-0}$ & ${3.845}^{+0.006}_{-0.005}$ & ${-13}^{+18}_{-15}$ \\
                6 & ${6.0}^{+0.3}_{-0.3}$ & ${0.55}^{+0.06}_{-0.06}$ & ${153}^{+30}_{-25}$ & ${3.887}^{+0.008}_{-0.007}$ & ${110}^{+24}_{-21}$ \\
                7 & ${7.8}^{+0.8}_{-1.1}$ & ${0.47}^{+0.10}_{-0.13}$ & ${137}^{+42}_{-38}$ & ${3.858}^{+0.011}_{-0.012}$ & ${25}^{+30}_{-35}$ \\
                8 & ${8.3}^{+0.6}_{-0.6}$ & ${0.46}^{+0.07}_{-0.07}$ & ${44}^{+54}_{-44}$ & ${3.881}^{+0.007}_{-0.012}$ & ${92}^{+19}_{-34}$ \\
                9 & ${6.5}^{+0.7}_{-0.6}$ & ${0.47}^{+0.11}_{-0.07}$ & ${60}^{+43}_{-60}$ & ${3.854}^{+0.009}_{-0.008}$ & ${14}^{+26}_{-22}$ \\
                10 & ${7.0}^{+0.7}_{-0.7}$ & ${0.46}^{+0.10}_{-0.10}$ & ${151}^{+49}_{-49}$ & ${3.893}^{+0.016}_{-0.013}$ & ${125}^{+45}_{-36}$ \\
                11 & ${7.0}^{+0.6}_{-0.6}$ & ${0.67}^{+0.11}_{-0.10}$ & ${254}^{+46}_{-41}$ & ${3.865}^{+0.017}_{-0.018}$ & ${45}^{+48}_{-50}$ \\
                12 & ${6.5}^{+0.7}_{-0.6}$ & ${0.55}^{+0.10}_{-0.09}$ & ${140}^{+46}_{-43}$ & ${3.908}^{+0.015}_{-0.014}$ & ${168}^{+42}_{-39}$ \\
                13 & ${6.7}^{+0.8}_{-0.7}$ & ${0.30}^{+0.08}_{-0.07}$ & ${146}^{+63}_{-62}$ & ${3.892}^{+0.019}_{-0.019}$ & ${122}^{+54}_{-55}$ \\
                \hline
                \multicolumn{6}{c}{$\mathrm{C\text{-}statistic} / \text{d.o.f : } 256965.3 / 441796$} \\
            \enddata
            \tablenotetext{\footnotesize a}{\footnotesize We adopted the proto-solar abundance table of \cite{Lodders2009}.}
            \tablenotetext{\footnotesize b}{\footnotesize The velocity reference is the BCG of A3571, with a redshift of $z=0.038583$ \citep{Smith2000}.
            A barycentric correction of $26.5 \, \mathrm{km \, s^{-1}}$ has been applied.}
        \end{deluxetable*}
   
        \begin{figure*}[ht!]
            \plotone{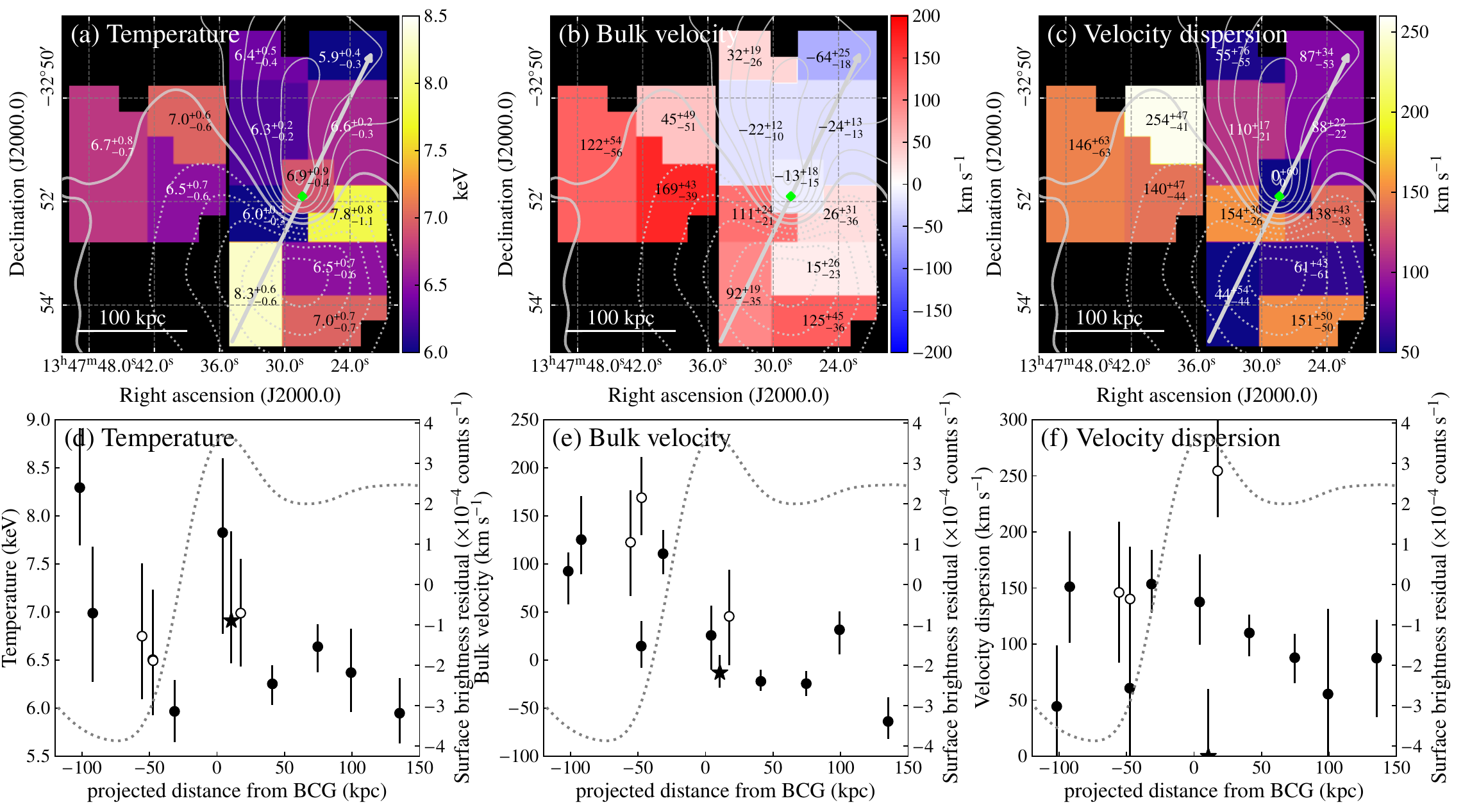}
            \caption{
                \textit{Upper panels:} Color maps of values of three spectral parameters: (a) temperature, (b) bulk velocity, and (c) velocity dispersion. 
                The gray contours represent the residuals from the elliptical $\beta$-model: thin solid lines indicate positive values, thick solid lines correspond to zero, and dotted lines denote negative values.
                The green diamond marks the position of the BCG,
                and the solid gray line indicates the projected axis passing through the BCG and ($13^\mathrm{h}47^\mathrm{m}24^\mathrm{s}$, \ang{-32;50;0.0}), which is used as the horizontal axis in the lower panels.
                \textit{Lower panels:} Panels (d), (e), and (f) show the temperature, bulk velocity, and velocity dispersion, respectively, as a function of projected distance from the BCG along the axis marked by the white line in the upper panels.
                Filled circles represent regions from the Center, North, and South observations, while open circles denote regions from the East observation.
                The gray dotted curve shows the XMM-Newton surface-brightness residual profile along the same projected axis.
                \label{fig:result}
            }
        \end{figure*}
 
        This model provides an acceptable fit to the data, with a C-statistic / d.o.f. = 256965/441796.
        Figure~\ref{fig:exspec} shows a representative spectrum and its best-fit SSM model for region 5. 
        The model reproduces the data well and no significant residuals are seen.
        Although this region contains the BCG, the fit does not require either an additional AGN-related power-law component or a multi-temperature ICM component.

    \subsection{Resolve Spectral Results}
        The best-fit parameters, $kT$, abundance, $\sigma_\mathrm{los}$, and redshift for each region are summarized in Table~\ref{tab:params}.
        The barycentric-corrected bulk velocity, $u_\mathrm{los}$, was defined relative to the BCG, ESO~383-G~076, 
        as $u_\mathrm{los} = c \times [(1+z_\mathrm{obs})(1+dz)/(1+z_\mathrm{BCG}) - 1]$.
        Here, we adopted $z_\mathrm{BCG} = 0.038583$ \citep{Smith2000} and 
        a barycentric correction of $\sim 26.5~\mathrm{km~s^{-1}}$ ($dz \sim 8.8 \times 10^{-5}$),
        evaluated at the midpoint of the observations (2025-01-06T07:07:34.000).
        Figure~\ref{fig:result} shows color maps and RA/Dec profiles of  $kT$, $\sigma_\mathrm{los}$, and $u_\mathrm{los}$.
       
        The BCG region has a temperature of $6.9^{+0.9}_{-0.4}$ keV. 
        The northern positive residual regions (regions 1--4) are cooler, at $\sim 5.9$--$6.6$ keV,
        while the southern negative-residual regions (regions 6--10) are hotter, at $\sim 6$--$8$ keV, with region 8 reaching $8.3 \pm 0.6$ keV.
        The eastern regions (regions 11--13) have temperatures of $\sim 6.5$--$7.0$ keV, comparable to that of the BCG region. 

        The bulk velocity roughly follows the surface-brightness residual structure, with the positive-residual regions showing blueshifts and the deficit regions showing redshifts.
        The relatively cooler gas in the northwestern regions is blueshifted, with the largest shift reaching $u_\mathrm{los}\sim -60 ~ \mathrm{km~s^{-1}}$, while the hotter gas in the southern and eastern regions is generally redshifted, reaching $u_\mathrm{los}\sim 70 ~ \mathrm{km~s^{-1}}$.
        The BCG region (region 5) shows a bulk velocity of $u_\mathrm{los}=-13 \pm 20 ~ \mathrm{km~s^{-1}}$, while 
        region~6, located immediately to the southeast, has $u_\mathrm{los}=110 \pm 20 ~ \mathrm{km~s^{-1}}$, implying a velocity difference of $\sim 130 ~ \mathrm{km~s^{-1}}$ between the two adjacent regions.
        The bulk velocity of region~11, is relatively small at $\sim 45 ~ \mathrm{km~s^{-1}}$, while the other eastern regions have $u_\mathrm{los}\sim 150 ~ \mathrm{km~s^{-1}}$.

        For region~5, which contains the BCG, $\sigma_\mathrm{los}$ is constrained to be $<60 ~ \mathrm{km~s^{-1}}$ at the 1 $\sigma$ level, suggesting that the gas there is dynamically quiet. 
        This value is lower than the velocity dispersion reported for the central region in \cite{McCall2026} ($115 \pm 15~\mathrm{km~s^{-1}}$). The difference is due to the different spatial binning schemes and region selections adopted in the two analyses. While the present work defines regions based on the residual image structure, \cite{McCall2026} uses bins designed to trace the X-ray emissivity distribution.
        The northern positive-residual regions (regions 1--4) also show relatively low velocity dispersion, in the range of $\sim 60$--$110~\mathrm{km~s^{-1}}$. 
        The southern deficit regions remain similarly quiet, with regions~8 and 9 having $\sigma_\mathrm{los} \lesssim 100~\mathrm{km~s^{-1}}$.
        By contrast, regions near the boundary between positive- and negative-residual structures show enhanced velocity dispersions: regions 6 and 7, south of the BCG have $\sigma_\mathrm{los}\sim 140\text{--}150 ~ \mathrm{km~s^{-1}}$.
        In particular, region~11, located \ang{;3;} ($\sim 150 ~ \mathrm{kpc}$) northeast of the BCG, lies near the boundary between the positive- and negative-residual structures and has almost no bulk velocity, shows the largest velocity dispersion, $\sigma_\mathrm{los}\sim 250^{+50}_{-40} ~ \mathrm{km~s^{-1}}$, 
        while the other eastern regions have values of $\sigma_\mathrm{los}\sim 140\text{--}150 ~ \mathrm{km~s^{-1}}$.

\section{Discussion}\label{sec:discussion}
    \subsection{A Sloshing Scenario}\label{subsec:sloshing}
        In A3571, we identify coherent gas motions that are spatially associated with the temperature structure.
        The northern cooler gas and the southern hotter gas are separated by a cold-front-like surface-brightness edge.
        The cooler gas (6.0--7.0~keV) in the northern region is blueshifted by $\sim 20 \text{--} 60 ~ \mathrm{km\,s^{-1}}$ relative to the BCG, whereas the hotter gas (6--8~keV) in the southern region is redshifted by $\sim 20 \text{--} 170 ~ \mathrm{km\,s^{-1}}$.
        Such a coherent velocity pattern is unlikely to be explained solely by random turbulent motions, instead indicating the presence of sloshing.

        However, there are no prominent spiral structures or multiple layers of cold fronts within $\sim200-300\,\rm{kpc}$, as are often found in other sloshing clusters.
        Instead, the X-ray surface brightness excess is elongated along the north–south direction.
        The cold-front-like edge lies south of the BCG,
        suggesting the relative motion with respect to the ambient atmosphere.
        We therefore speculate that the system is still in an early stage of sloshing development, with the low-entropy gas core undergoing its first oscillation within the gravitational potential (see a numerical demonstration in Section~\ref{subsec:purterber}).

        The velocity dispersion is largely uniform at $\simeq 50$–$150\, \mathrm{km\,s^{-1}}$ in our measurements, with an enhancement only in Region 11 ($\simeq 250\, \mathrm{km\,s^{-1}}$).

        \begin{figure*}[ht!]
            \plotone{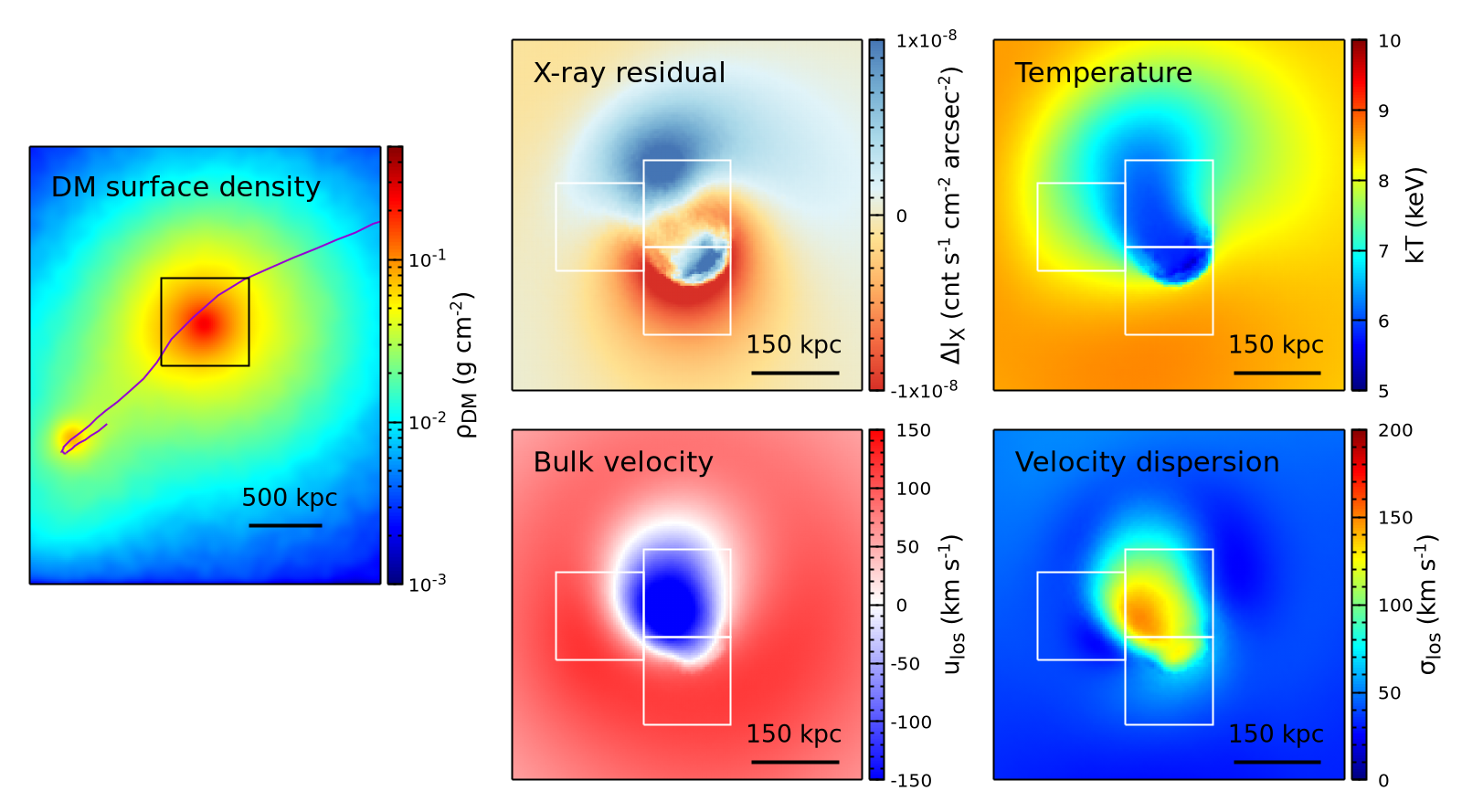}
            \caption{
                \textit{Left panel:} Simulated dark matter (DM) surface mass density distribution, which resembles the relative positions of A3571 (the main cluster) and A3572 (the subcluster). The purple line indicates the subcluster's infalling trajectory. \textit{Right panels:} Simulated X-ray surface brightness residual, X-ray-weighted temperature, X-ray-weighted line-of-sight bulk velocity, and velocity dispersion, zoomed into the region indicated by the black box marked in the left panel. The white squares approximately indicate the regions of the XRISM measurements (see Fig.~\ref{fig:result}).
                \label{fig:simulation}
            }
        \end{figure*}
        
    \subsection{A Perturber Candidate}\label{subsec:purterber}
        A3571 is located at the edge of the Shapley supercluster.
        Galaxy surveys indicate that the large-scale structure extends several Mpc along multiple directions (e.g., north, northwest, and south) from the cluster \citep{Haines2018,Baier-Soto2025},
        suggesting plausible infall trajectories for a merging subcluster.

        Southeast of A3571,
        at a distance of $\sim \ang{;30;}$ ($\sim 1.6 ~ \mathrm{Mpc}$), lies the optically-identified galaxy concentration Abell~3572 (A3572, $z=0.039$; BCG: $z=0.041$).
        The BCG of A3572 is redshifted by $\sim 640 ~\mathrm{km~s^{-1}}$ relative to that of A3571 \citep{Lauer2014}. 
        Although A3572 contains on the order of $\sim$ 50 optical members \citep{Abell1989},
        observations with Einstein Probe have not detected significant X-ray emission at its position \citep{Zheng2026}.
        A3572 is a promising perturber candidate for the system and may have lost most of its gaseous atmosphere while moving within the supercluster environment before colliding with A3571.

        We tested this scenario using idealized cluster merger simulations,
        widely employed to model individual merging clusters and interpret their multi-wavelength observations \citep[e.g.,][]{Springel2007,Sheardown2019},
        including recent XRISM targets \citep[e.g.,][]{Heinrich2025,Zhang2026,Bellomi2025}.
        The simulations, performed with the moving-mesh code Arepo \citep{Springel2010,Weinberger2020},
        model a merger between two initially spherical galaxy clusters.
        Each cluster contains a dark matter halo with a radial density profile following the Navarro–Frenk–White (NFW) form (see \citealt{Zhang2014} for more details of our numerical set-ups).
        The gas density radial profile of the main cluster follows the best fit to A3571 based on Chandra observations.
        The radial temperature profile is estimated under the assumption of hydrostatic equilibrium.
        For simplicity, we assume that the subcluster is gasless, motivated by its lack of X-ray emission.
        We fix the mass of the main cluster ($M_\mathrm{vir} = 8\times10^{14}\,M_\odot$) in all our simulations \citep{Nevalainen2001}.
        We also fix the initial pairwise velocity ($V_0 = 500{\,\rm km\,s^{-1}}$) between the two merging clusters, and explore other parameters that determine the merging halo trajectories, including the merger mass ratio ($\xi=4-16$) and impact parameter ($P_0=0-1\,{\rm Mpc}$).
    
        Fig.~\ref{fig:simulation} shows an example of our simulations that can broadly capture the key observational features of A3571.
        {The X-ray residual is estimated using the same approach as for the observational image (see Fig.~\ref{fig:A3571img}).
        The snapshot corresponds to $\sim1~\mathrm{Gyr}$ after the primary pericentric passage of a minor, off-axis merger with a merger mass ratio of 8:1 and an impact parameter of $1\,{\rm Mpc}$.
        The merging subcluster is approaching its apocenter with a relative velocity of $\sim400~\mathrm{km~s^{-1}}$ to the main cluster along the line of sight.
        As the subcluster passes by, the cold, dense gas core of the main cluster is gravitationally perturbed, producing the observed elongated X-ray morphology.
        Our viewing direction is inclined by $\sim60^{\circ}-70^{\circ}$ with respect to the normal to the merger plane.
        The bulk velocity distribution traces the motion of the gas core relative to the surrounding atmospheres.
        Rayleigh–Taylor instabilities will soon be triggered near the cold front (in a pure hydrodynamic situation), followed by the formation of a sloshing spiral structure in the next few Gyrs.
        All of these features generally reproduce what XRISM has observed.
        In our scenario, A3571 is in an early stage of the sloshing development, similar to the case of the Coma cluster \citep{Zhang2026_Coma}.
        We note that our simulation predicts a slightly higher velocity dispersion near the edge of the perturbed core, which, however, is not as significant as that observed in Region 11.
        It is still unclear if the measured high velocity dispersion ($\simeq250~\mathrm{km~s^{-1}}$) can be explained by a minor-merger scenario.}
    
        We emphasize that the goal of our simulations is to identify a plausible merger configuration for A3571 rather than to fine-tune all its observational details.
        The simulated gas distributions are likely affected by the cluster's intrinsic ellipticity, initial radial gas profiles, and other factors.
        In particular, A3571 hosts one of the most massive BCGs, which exhibits an extended optical distribution (with a diameter of $\sim300\,~\mathrm{kpc}$) and high ellipticity \citep{Kemp1991}.
        The shape of the BCG is often a robust tracer of the underlying dark matter halo \citep[e.g.,][]{1982A&A...107..338B,2019MNRAS.490.4889H}.
        The resulting elliptical gravitational potential may therefore affect the detailed gas distribution, which cannot be captured by our idealized simulations.
        Moreover, the alignment between the BCG shape and the elongated X-ray morphology may indicate that continuous accretion, including the merging of small substructures, along preferred filamentary directions contributes to shaping the elongated distributions of dark matter, gas, and stars \citep[e.g.,][]{2005ApJ...627..647B}. 

    \subsection{Implications for the Absence of a Cool Core in A3571}\label{disrupt}
        A3571 presents an intriguing case in which the global X-ray morphology is relatively regular and the ICM velocity dispersion is low, yet the cluster does not host a well-developed cool core.
        In relaxed cool-core clusters such as Perseus, Centaurus, Hydra~A, and Abell~2029, XRISM has measured velocity dispersions of typically
        $\sim 50\text{--}200~\mathrm{km~s^{-1}}$, indicating predominantly subsonic ICM motions
        \citep{XRISM2025_per, XRISM2025_cen, XRISM2025_A2029, XRISM2026_M87, Tanaka2026, Rose2025}.
        A3571 shows a comparable level of line broadening, with mainly 
        $\sim 50\text{--}200~\mathrm{km~s^{-1}}$ velocity dispersions and an enhancement to $\sim 250~\mathrm{km~s^{-1}}$ only in Region~11.
        This is distinct from systems more dynamically disturbed such as Abell~2319 and Abell~754, where larger velocity dispersions have been reported
        \citep{XRISM2025_A2319, 
        Omiya2026_A754, Omiya2026_A3667, Fujita2025, Heinrich2025, Ota2026, Heinrich2026}.
        Thus, the absence of a cool core in A3571 is unlikely to be attributed to widespread strong turbulent broadening driven by a major merger.

        The central cooling time of A3571, $\sim 2.1~\mathrm{Gyr}$ \citep{Hudson2010}, is longer than is typically observed in strongly developed cool cores.
        This relatively long cooling time is consistent with the absence of a compact, rapidly cooling core.
        The coherent velocity pattern and the association between the cooler gas and the positive X-ray surface-brightness residuals suggest a milder process: merger-induced sloshing may have displaced the low-entropy gas from the potential center and reduced its apparent concentration.

        The plausible merger configuration inferred from our simulations corresponds to a stage $\sim 1~\mathrm{Gyr}$ after the first pericentric passage, which is shorter than the current central cooling time.
        Radiative cooling may therefore not yet have had sufficient time to establish, or re-establish, a centrally concentrated cool core after the perturbation.
        In relaxed cool-core systems, mechanical AGN feedback is thought to offset radiative cooling, and XRISM has revealed localized velocity-dispersion enhancements around central AGNs in some systems \citep{XRISM2025_per, XRISM2026_M87}.
        In A3571, however, sloshing may have displaced dense, low-entropy gas away from the BCG, thereby weakening the AGN feedback cycle.
        If the central AGN is fueled by gas cooling or accumulating in the innermost core, the reduced central gas reservoir would naturally limit the fuel supply and lower the mechanical heating power.
        This possible connection between merger-driven gas redistribution and AGN fueling/feedback provides a qualitative link to recent cosmological simulations of cool-core/non-cool-core transformations, in which both merger-driven perturbations and AGN feedback may play important roles \citep{Lehle2025}.

        A3571 may therefore represent a regular, low-turbulence cluster in which a developed cool core is absent because merger-induced sloshing has reduced the central concentration of low-entropy gas.
        A3571 thus provides a direct XRISM view of coherent sloshing in a morphologically regular cluster, highlighting merger-driven gas redistribution, rather than strong turbulence, as a key process shaping the absence of a developed cool core.

\begin{acknowledgments}
    This work was supported by JST SPRING, Grant Number JPMJSP2151.
    CZ acknowledges the support of the Czech Science Foundation (GACR) Junior Star grant no.~GM24-10599M.
    HM and IZ were partially supported by NASA grant 80NSSC25K7662.
    The simulations presented in this paper were carried out using the Midway computing cluster provided by the University of Chicago Research Computing Center.
    WF and CJ acknowledge support from the Smithsonian Institution, the Chandra High Resolution Camera Project through NASA contract NAS8-0306.
    IK was supported by the Simons Foundation via the Simons Investigator Award to A. A. Schekochihin.
\end{acknowledgments}

\bibliography{ApJL}
\bibliographystyle{ApJL}



\end{document}